%% file: article.tex
\documentclass[12pt]{spieman}  
\usepackage{amsmath,amsfonts,amssymb}
\usepackage{graphicx}
\usepackage{setspace}
\usepackage{tocloft}
\usepackage{array,makecell}

\title{Deep Learning Assisted Modeling for $\chi^{(2)}$ Nonlinear Optics}

\author[a,b,*]{Jack Hirschman}
\author[c]{Erfan Abedi}
\author[d,e]{Minyang Wang}
\author[b,c]{Hao Zhang}
\author[c]{Abhimanyu Borthakur}
\author[d]{Justin Baker}
\author[d]{Andrea L. Bertozzi}
\author[b]{Randy Lemons}
\author[b,c,f,g]{Sergio Carbajo}
\affil[a]{Stanford University, Applied Physics, 348 Via Pueblo, Stanford, CA 94305, USA}
\affil[b]{SLAC National Accelerator Laboratory, 2575 Sand Hill Rd, Menlo Park, CA 94025, USA}
\affil[c]{University of California, Los Angeles, Electrical \& Computer Engineering, 420 Westwood Plaza, Los Angeles, CA 90095, USA}
\affil[d]{University of California, Los Angeles, Mathematics, 520 Portola Plaza, Los Angeles, CA 90095, USA}
\affil[e]{University of California, Los Angeles, Statistics \& Data Science, 520 Portola Plaza, Los Angeles, CA 90095, USA}
\affil[f]{University of California, Los Angeles, Physics \& Astronomy, 475 Portola Plaza, Los Angeles, CA 90095, USA}
\affil[g]{California NanoSystems Insitute, 570 Westwood Plaza, Los Angeles, CA 90095, USA}

\cftpagenumbersoff{figure}
\cftpagenumbersoff{table} 
\begin{document} 
\maketitle

\begin{abstract}
Modeling second-order ($\chi^{(2)}$) nonlinear optical processes remains computationally expensive due to the need to resolve fast field oscillations and simulate wave propagation using methods like the split-step Fourier method (SSFM). 
This can become a bottleneck in real-time applications, such as high-repetition-rate laser systems requiring rapid feedback and control.
We present an LSTM-based surrogate model trained on SSFM simulations generated from a start-to-end model of the photocathode drive laser at SLAC National Accelerator Laboratory’s Linac Coherent Light Source II. 
The model achieves over 250× speedup while maintaining high fidelity, enabling future real-time optimization and laying the foundation for data-integrated modeling frameworks and digital twins of laser systems.
 
\end{abstract}


{\noindent \footnotesize\textbf{*}Jack Hirschman,  \linkable{jhirschm@stanford.edu} }

\begin{spacing}{1}   

\section{Introduction}
\label{sect:intro}  
\input{sections/intro}

\section{Results}
\label{sect:intro}  
\input{sections/results}

\section{Discussion}
\label{sect:intro}  
\input{sections/discussion}

\section{Conclusions}
\label{sect:intro}  
\input{sections/conclusions}

\subsection*{Acknowledgments}
 The authors acknowledge the support from the SLAC National Accelerator Laboratory, the U.S. Department of Energy (DOE), the Office of Science, Office of Basic Energy Sciences under Contract No. DE-AC02-76SF00515, No. DE-SC0022559, No. DE-FOA-0002859, the National Science Foundation under Contract No. 2231334 and 2436343, and the U.S. Department of Defense via AFOSR Contract No. FA9550-23-1-0409 and the National Defense Science and Engineering Graduate Fellowship.

 Furthermore, the authors acknowledge the use of large language models, specifically ChatGPT, for language and grammar clean-up. 
 Additionally, the authors would like to acknowledge Gregory Stewart for his help with graphic editing.

\subsection*{Disclosures}
The authors declare that there are no financial interests, commercial affiliations, or other potential conflicts of interest that could have influenced the objectivity of this research or the writing of this paper.

\subsection* {Code, Data, and Materials Availability} 
Data supporting this study can be found on the Stanford Digital Repository at \\\hyperlink{https://doi.org/10.25740/nf288ry2198}{https://doi.org/10.25740/nf288ry2198}.
Additionally, the code is made publicly available on GitHub at \hyperlink{https://github.com/jhirschm/DCNS_LSTM_Public}{https://github.com/jhirschm/DCNS\_LSTM\_Public}.
Please see the Supplementary Information for more details or reach out to the corresponding author with reasonable requests.

\bibliography{report}   
\bibliographystyle{spiejour}   



\end{spacing}
\end{document}

%% file: sections/intro.tex
Nonlinear optical processes governed by second-order ($\chi^{(2)}$) interactions play a fundamental role in modern photonics, enabling key functionalities in quantum information science, integrated photonic circuits, biomedical imaging, and high-power laser systems. 
These interactions, including sum-frequency generation (SFG), second-harmonic generation (SHG), and difference-frequency generation (DFG), underpin technologies such as entangled photon pair generation~\cite{kultavewuti2017polarization, zhang2021high}, quantum frequency conversion~\cite{mckenna2022ultra}, and high-speed electro-optic modulation~\cite{luo2019nonlinear}.
In imaging, $\chi^{(2)}$ interactions facilitate label-free microscopy techniques like SHG microscopy for visualizing biological structures~\cite{campagnola2003second} and SFG spectroscopy for surface analysis~\cite{shen1989surface}.

In high-power laser applications, $\chi^{(2)}$ nonlinearities are essential for optical parametric amplification (OPA), which generates ultrafast pulses with tunable wavelengths, widely used in high-intensity lasers~\cite{cerullo2003ultrafast}. 
In accelerator physics, these nonlinear processes play a critical role in photocathode drive lasers, where precise frequency conversion and pulse shaping are essential for generating high-brightness electron beams~\cite{zhang2024linac}. 
These lasers rely on precise pulse shaping and frequency conversion, even utilizing noncollinear phase-matching in $\chi^{(2)}$ crystals to optimize electron emission and beam quality.

While $\chi^{(2)}$ processes are well-understood theoretically, their numerical modeling remains computationally expensive due to the need to resolve fast field oscillations for Fourier-based methods and to simulate discrete propagation through media using techniques like the split-step Fourier method (SSFM)~\cite{couairon2011practitioner, yee1966numerical, butcher2015runge}. 
These demands are particularly taxing in real-time contexts, where models must operate fast enough to interface with high-repetition-rate experiments or streaming diagnostics.
In such cases, the computational burden can become a practical bottleneck, limiting the deployment of adaptive control or optimization strategies. 
Furthermore, for discovery science applications—where experimental imperfections, noise, or system drift are critical to capture—traditional physics-based models often fall short, as they lack the flexibility to integrate real-world variability.
This creates an urgent need for modeling frameworks that are both fast and robust to imperfections, enabling real-time feedback, exploration of complex nonlinear optical phenomena, and inverse design. 
The need for real-time modeling and integration with experiment becomes even more important with the greater push towards digital twin development in larger facilities that rely on cascaded simulations involving nonlinear ~\cite{edelen2024digital, edelen2019machine, biedron2022snowmass21}.

The increasing demand for real-time modeling and predictive optimization has fueled interest in machine learning (ML) as an alternative computational framework~\cite{salmela2021predicting,salmela2022feed,sui2023predicting,lauria2024conditional,pu2023fast,jiang2023predicting,zhang2025multi}. 
Recurrent neural networks (RNNs), particularly Long Short-Term Memory (LSTM) networks, have demonstrated their capability in capturing the temporal dynamics of nonlinear optical systems~\cite{salmela2021predicting}. 
By adopting ML surrogates, simulations can be dramatically accelerated while maintaining high accuracy, enabling real-time optimization and control.

In this work, we present an LSTM-based surrogate model that dramatically reduces computational cost while accurately predicting the output fields of noncollinear SFG. 
This configuration serves as a canonical example of $\chi^{(2)}$ nonlinear optics, characterized by strong sensitivity to pulse structure, nontrivial phase-matching conditions, and the interplay of spatial and temporal effects. Noncollinear SFG is widely used in ultrafast optics and nonlinear frequency conversion systems, making it a broadly relevant testbed for assessing surrogate model fidelity and generalizability. 
Training data is generated using high-resolution SSFM~\cite{hirschman2024design}, enabling the model to learn the coupled spatiotemporal dynamics that govern nonlinear field evolution.
We evaluate the surrogate in terms of accuracy, inference speed, and its capacity to support real-time optimization and adaptive laser control.

Beyond serving as a stand-alone replacement for traditional solvers, this work lays the foundation for inverse design methodologies and digital twin development in nonlinear optics~\cite{hirschman2024design, hirschman:ipac2025-mopb040}. 
By bridging high-fidelity simulation with real-time predictive modeling, ML-based surrogates enable scalable optimization, facilitate integration with experimental diagnostics, and provide a pathway toward self-updating, hybrid models that incorporate both physics and data. 
These capabilities are critical for advancing closed-loop control, scalable simulation workflows, and next-generation photonic system design.

%% file: sections/results.tex
\subsection{Model \& Metrics}
\begin{figure}
\centering
\includegraphics[width=\textwidth]{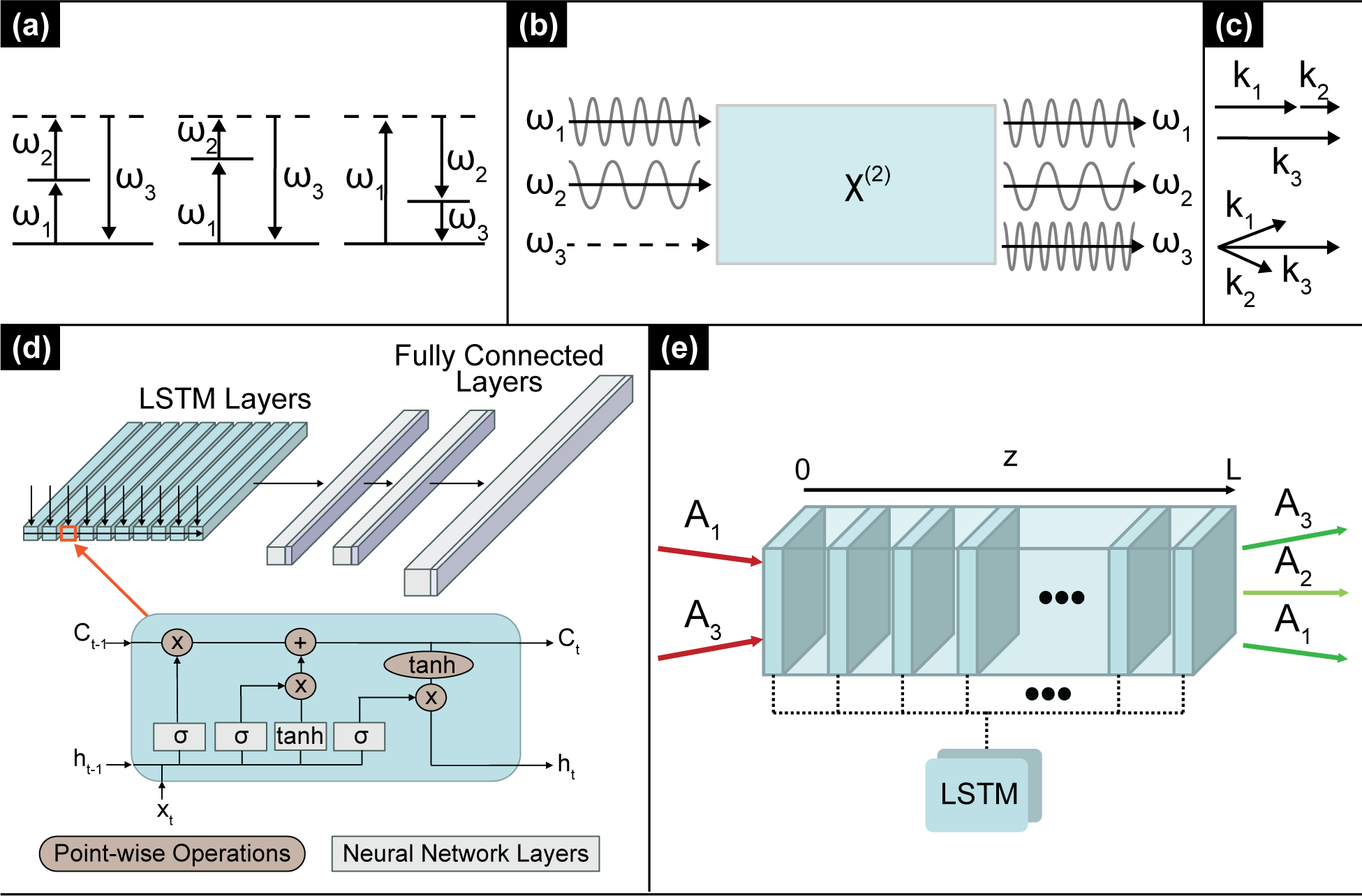}
\caption 
{ \label{fig:main_fig}
$\chi^{(2)}$ Processes and Neural Network Architecture. a) schematic of SHG, SFG, and DFG $\chi^{(2)}$ processes; b) input-output block diagram of $\chi^{(2)}$ nonlinear medium; c) associated wave vectors for $\chi^{(2)}$ processes; d) single-layer 10-input LSTM architecture with output fully connected layers and internal workings of LSTM block; and e) noncollinear SFG process with input and output fields as well as encapsulated LSTM network for discretized replacement.} 
\end{figure}

The simulation of noncollinear SFG in nonlinear optical media requires solving the Generalized Nonlinear Schr\"odinger Equation (GNSLE)~\cite{agrawal2012nonlinear,brabec1997nonlinear,lemons2022temporal}, a set of coupled differential equations describing the interaction between the three optical  is given by (Eq.~(\ref{eq:a1})--(\ref{eq:a3}) and Fig.~\ref{fig:main_fig}a–c)
\begin{equation}
\label{eq:a1}
\frac{dA_{1}}{dz} = \frac{2 i d_{\mathrm{eff}} \, \omega}{k_{1}^{2} c} \, A_{2} \, A_{3} \, e^{-i \Delta k z}
\end{equation}

\begin{equation}
\label{eq:a2}
\frac{dA_{2}}{dz} = \frac{2 i d_{\mathrm{eff}} \, \omega}{k_{2}^{2} c} \, A_{1} \, A_{3} \, e^{-i \Delta k z}
\end{equation}

\begin{equation}
\label{eq:a3}
\frac{dA_{3}}{dz} = \frac{2 i d_{\mathrm{eff}} \, \omega}{k_{3}^{2} c} \, A_{1} \, A_{2} \, e^{-i \Delta k z},
\end{equation}

where z is the propagation axis, $d_{\mathrm{eff}}$ is the effective nonlinearity, $A_1$--$A_3$ are the interacting fields, $\omega$ is the angular frequency, and $k$ represents the phase mismatch (see Fig.~\ref{fig:main_fig}). 

\begin{figure}
\centering
\includegraphics[width=\textwidth]{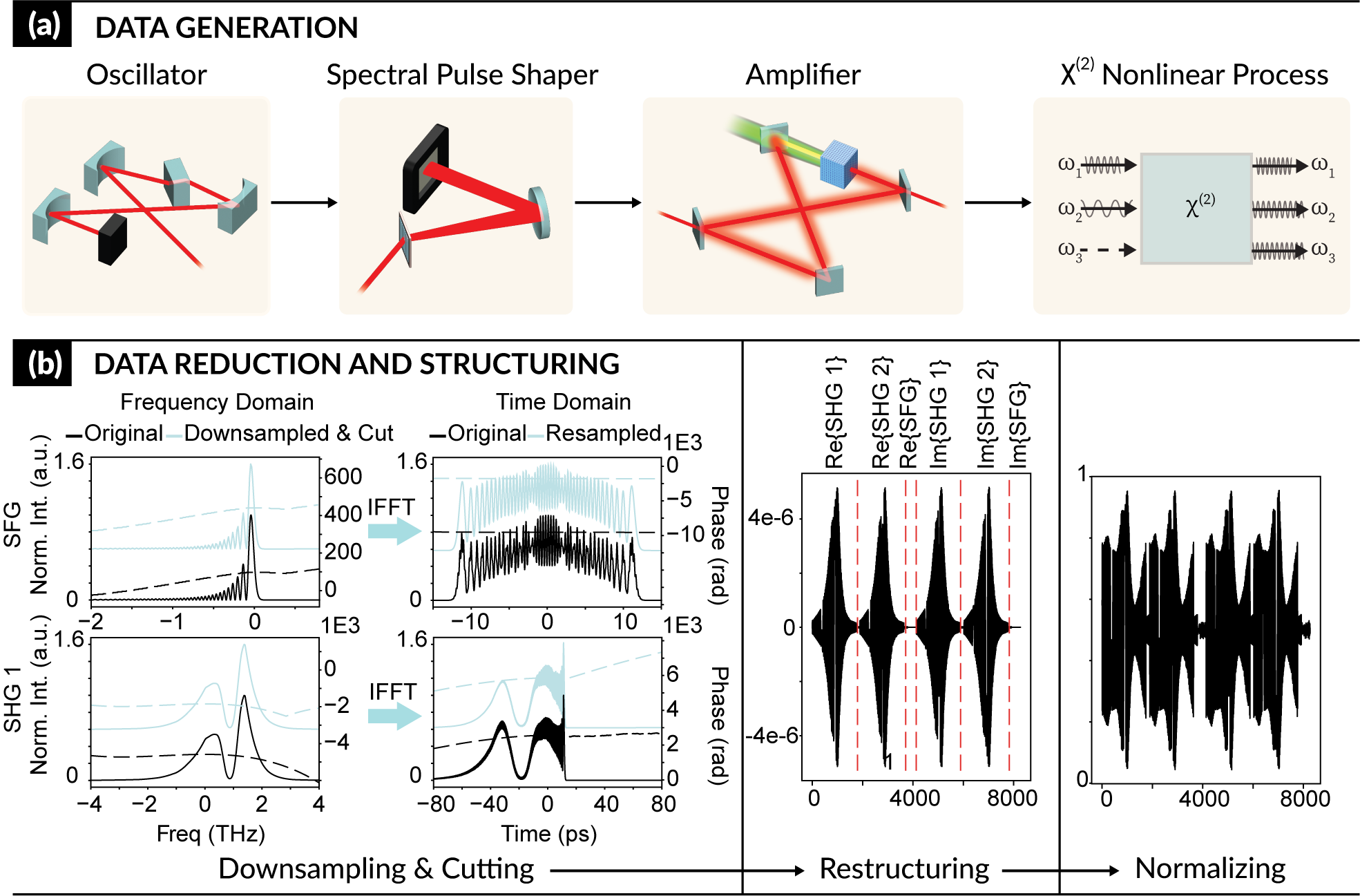}
\caption 
{ \label{fig:data_proc}
Data Generation and Preparation for LSTM. a) start-to-end model of laser system~\cite{hirschman2024design} used to generate the data and b) the three-stage processing, including downsampling and cutting, restructuring into one vector, and normalization.} 
\end{figure}

Solving these equations using SSFM can present a relatively significant computational cost. For example, in the full start-to-end simulation of our laser system (see Fig.~\ref{fig:data_proc}a), SSFM-based calculations for this nonlinear upconversion cover roughly 95\% of the entire simulation time and require tracking three highly-sampled fields in time and frequency for each step in the nonlinear medium.
We, thus, implement an LSTM-based surrogate model to predict field evolution efficiently (Fig.~\ref{fig:main_fig}e). 
The LSTM is structured as a sequence-to-sequence model, where each discretized slice of the nonlinear crystal is treated as a time step. 
The network consists of 2048 hidden units, followed by three fully connected layers with dimensions (2048, 4096), (4096, 4096), and (4096, 8624), utilizing ReLU, Tanh, and Sigmoid activations, respectively (Fig.~\ref{fig:main_fig}d). 

To provide the LSTM with sufficient memory while avoiding excessive reliance on past inputs, we use a history of 10 spatial slices, determined through empirical tuning to balance performance and latency. 
The network is trained on 890,000 instances, validated on 10,000 instances, and tested on 90,000 instances. 
Optimization is performed using the Adam optimizer and a weighted mean squared error (wMSE) loss function:
\begin{equation}
\label{eq:wmse}
wMSE = c_{\mathrm{SFG}} \cdot MSE_{\mathrm{SFG}(A_{2})} 
      + c_{\mathrm{SHG}} \cdot \Bigl( MSE_{\mathrm{SHG1}(A_{1})} 
      + MSE_{\mathrm{SHG2}(A_{3})} \Bigr).
\end{equation}

The $c_{\mathrm{SFG}}$ weight corresponds to $A_{2}$ and the  $c_{\mathrm{SHG}}$ weight corresponds to fields $A_{1}$ and $A_{3}$.
These weights were empirically chosen as 0.7 and 0.3, respectively, to rescale for emphasis on the SFG field, which has a smaller contribution to the total concatenated vector as represented in Fig.~\ref{fig:data_proc}b. 

To evaluate model inference performance, we employ a composite error metric designed to balance sensitivity to waveform shape and total energy.
Temporal intensity profiles are obtained by taking the squared magnitude of the inverse Fourier transform of the complex fields. 
These profiles are cropped to shaping-relevant windows--SFG ($A_{2}$): -10.1 to 10.1 ps; SHG1 ($A_{1}$): -75 to 25 ps; SHG2 ($A_{3}$): -25 to 75 ps--to focus the comparison on the most salient temporal regions. 
Gaussian smoothing is applied to individual metric components to suppress high-frequency numerical artifacts without degrading large-scale structure.

The composite metric consists of three unitless components: (i) Cosine similarity (CosSim) of area-normalized waveforms~\cite{irimatsugawa2021cosine}, inverted and scaled so that identical profiles yield zero error; (ii) energy-proportional normalized mean squared error (NMSE) computed from the total integrated intensity, penalizing mismatches in energy ; and (iii) Wasserstein distance (Earth Mover’s Distance, EMD) between the two intensity profiles, sensitive to localized redistributions of intensity~\cite{frogner2015learning}.

Each component is normalized and mapped to [0, 1] using fixed clipping ranges, with lower values corresponding to better agreement. 
The final score is computed via quadratic combination:
\begin{equation}
\label{eq:eval}
M_{\mathrm{combined}} =
\sqrt{ \, 
    w_{\mathrm{CosSim}} \cdot s_{\mathrm{CosSim}}^{2}
  + w_{\mathrm{NMSE}}   \cdot s_{\mathrm{NMSE}}^{2}
  + w_{\mathrm{EMD}}    \cdot s_{\mathrm{EMD}}^{2}
},
\end{equation}
where $w_{\mathrm{CosSim}}$ = 0.4, $w_{\mathrm{NMSE}}$ = 0.5, $w_{\mathrm{EMD}}$ = 0.1, and $s_{\mathrm{i}}$ denotes the associated scaled error.
Lower values of $M_{\mathrm{combined}}$ indicate closer agreement between predicted and reference profiles.

This formulation was selected after early testing showed that single metrics such as NMSE of field or intensity profile, cosine similarity, or structural similarity index measure (SSIM) alone often failed to align with qualitative visual assessments across diverse pulse shapes. 
By combining complementary terms sensitive to both structural and energetic fidelity and empirically tuning the hyperparameters, the metric produces rankings that better reflect perceived agreement in the temporal domain (see Supplementary Information).

\subsection{Data Generation}
Figure~\ref{fig:data_proc} outlines the data generation and preprocessing steps. 
Simulation data is produced using a start-to-end model of the SLAC Linac Coherent Light Source II (LCLS-II) photoinjector laser system~\cite{hirschman2024design,zhang2024linac}, which includes a 1035 nm mode-locked infrared laser source, a programmable spectral amplitude and phase shaper, a regenerative chirped pulse amplifier, and a nonlinear frequency conversion stage (see Fig.~\ref{fig:data_proc}a). 
The final ultraviolet pulses are generated via dispersion-controlled nonlinear synthesis (DCNS), a process in which the noncollinear sum-frequency generation (SFG) of temporally dispersed IR pulses enables control over the resulting temporal profile~\cite{lemons2022temporal}.
By tuning second- and third-order dispersion (SOD and TOD), DCNS facilitates tailored UV pulse shaping critical for photocathode emission optimization. 
To capture a broad range of profiles, we generate 10,000 pulse shaper configurations by randomly sampling SOD, TOD, and amplitude shaping parameters (Table~\ref{tab:data_params}) for the upstream pulse shaper, with at least 400 cases employing phase-only shaping. 
These selected combinations generate pulses fed to the noncollinear SFG process to build the data set.

\begin{table}[h!]
\centering
\caption{\label{tab:data_params}Parameter limits for data generation. See Hirschman~\textit{et al.}~\cite{hirschman2024design} for details on pulse shaping simulation.}
\renewcommand{\arraystretch}{1}
\begin{tabular}{|>{\centering\arraybackslash}m{1cm}|
                >{\centering\arraybackslash}m{2cm}|
                >{\centering\arraybackslash}m{2cm}|
                >{\centering\arraybackslash}m{2.4cm}|
                >{\centering\arraybackslash}m{2.5cm}|
                >{\centering\arraybackslash}m{2.5cm}|}
\hline
 & \textbf{SOD (fs$^{2}$)} & \textbf{TOD (fs$^{3}$)} & \textbf{Hole Position (nm)} & \textbf{Hole Depth (\%)} & \textbf{Hole Width (nm)} \\ \hline
\textbf{Min} & -1E4 & -1E5 & 1022 & 0 & 0.1 \\ \hline
\textbf{Max} &  1E4 &  1E5 & 1036 & 95 & 4   \\ \hline
\end{tabular}
\end{table}

Each configuration is simulated using the split-step Fourier method (SSFM), yielding the complex-valued optical fields $A_{1}$, $A_{2}$, and $A_{3}$--corresponding to SHG1, SFG, and SHG2--at each of 100 discretized slices through the nonlinear crystal. 
Each field is sampled with 32,768 complex points to resolve carrier-scale dynamics and frequency-dependent phase evolution. 
To reduce computational burden and enable efficient surrogate modeling, we apply a structured three-stage preprocessing workflow to the simulated fields, illustrated in Fig.~\ref{fig:data_proc}b.

In the first stage, each field is truncated and downsampled in the frequency domain.
This choice reflects the relative invariance of spectral bandwidth to pulse duration compared to the time domain, making the frequency representation more consistent across varying dispersion conditions. 
The SFG field ($A_{2}$) is reduced to 348 complex elements, while SHG1 and SHG2 fields ($A_{1}$ and $A_{3}$) are reduced to 1,892 complex elements each. 
We validate this reduction by interpolating the spectra back to the original resolution and reconstructing time-domain intensity profiles via inverse FFT, confirming agreement with the full-resolution simulation.

In the second stage, the real and imaginary parts of each field are concatenated to form a single real-valued vector. 
The ordering consists of all real components ($A_{1}$, $A_{2}$, $A_{3}$) followed by the corresponding imaginary components, resulting in a fixed-length vector of 8,264 elements. 
In the third stage, each element is normalized to the range [0, 1] based on the global dataset-wide extrema to ensure consistent scaling across all training instances.

After preprocessing, the data is formatted for supervised training of the LSTM model. 
The LSTM receives a sequence of 10 spatial slices as input to predict the subsequent slice (see Supplementary Information). 
Each simulation of 100 slices yields 100 input-output pairs by applying a sliding window across the sequence. 
For the first nine instances, the input sequence is prepended with repeated copies of the initial slice to maintain consistent input length. 
The resulting training tensor shapes are (batch size, 10, 8264) for the input X and (batch size, 8264) for the output y.
During inference, the model operates autoregressively. 
The first input is formed by repeating the initial slice ten times. The LSTM then predicts the next slice, which is appended to the input sequence, and the oldest slice is discarded to maintain a window of length ten. 
This process is repeated for all 100 spatial steps, effectively reconstructing the entire field evolution through the nonlinear crystal using sequential model predictions.

\subsection{Reconstruction}

The LSTM converges over about 180 epochs, performed on a single NVIDIA A10G GPU. 
The final training and validation losses reached $2.05\times10^{-5}$ and $2.03\times10^{-5}$, respectively, after approximately 160 hours. 
During inference, the model is run sequentially for each of the 100 spatial steps with each prediction dependent on the preceding ten slices.

\begin{figure}
\centering
\includegraphics[width=\textwidth]{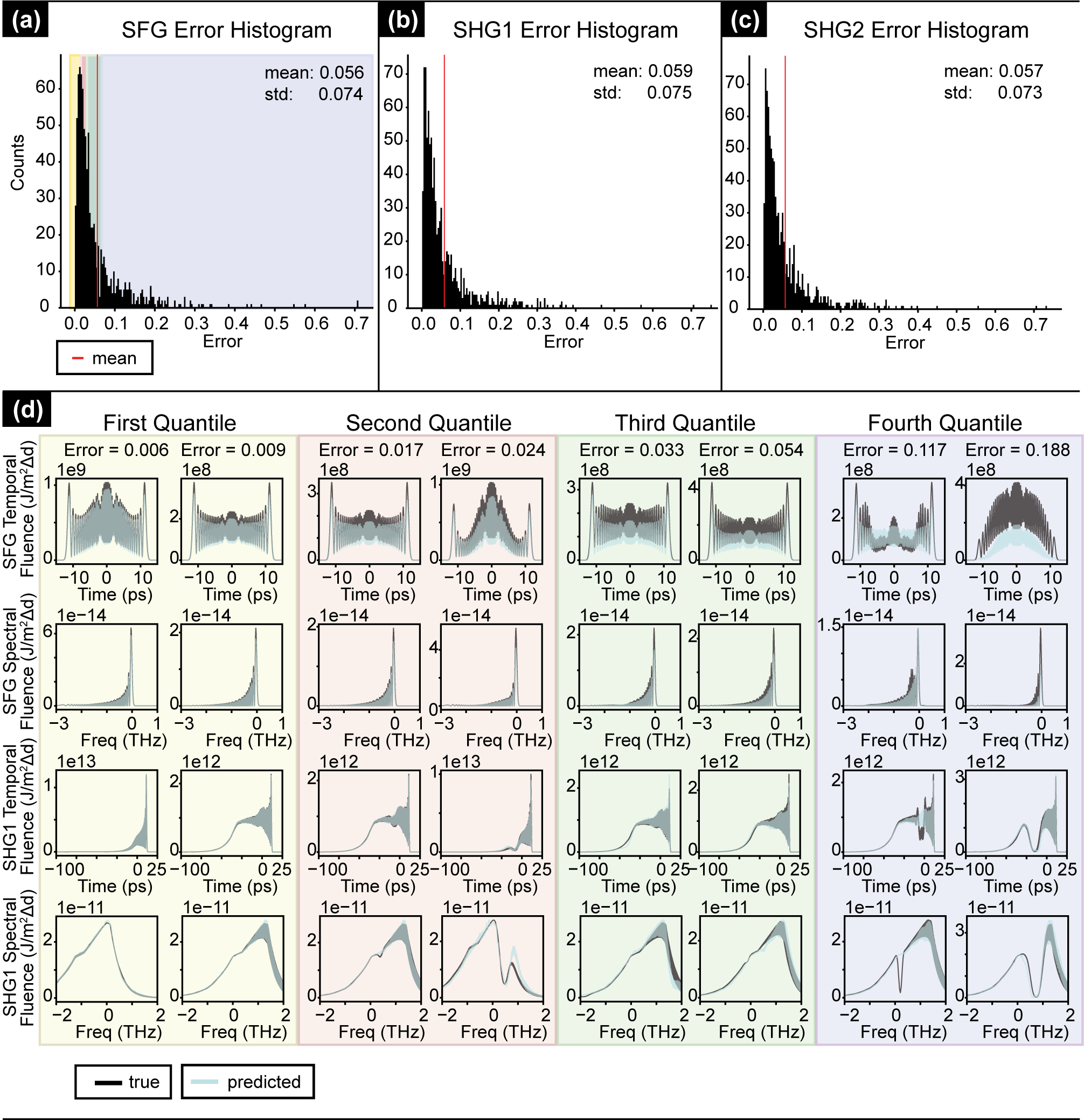}
\caption 
{ \label{fig:err_hist}
Evaluation Error Distribution. Histograms and statistics for the combined error metric of temporal intensity for a) SFG, b) SHG1, and c) SHG2, respectively, and d) two randomly selected examples from the test data set from within each error quartile of the SFG error distribution (shaded) accompanied by the temporal and spectral predicted and true intensity profiles for the SFG signal and the corresponding SHG1 signal.
} 
\end{figure}

To assess reconstruction fidelity, we evaluate the combined error metric between the predicted and SSFM-generated outputs. 
Figure~\ref{fig:err_hist}a--c presents the error distributions for the temporal intensity profiles of the SFG ($A_{2}$), SHG1 ($A_{1}$), and SHG2 ($A_{3}$) signals. 
The SFG output serves as the primary performance target given its critical role in UV pulse shaping and its dynamic nature during nonlinear propagation. 
Unlike the SHG signals, which evolve relatively gradually through the crystal, the SFG field is initially absent and generated entirely via interaction, undergoing significant spectral and temporal reshaping. 
As such, it provides a more stringent and informative test of model fidelity. 
To visualize qualitative performance, Fig.~\ref{fig:err_hist}d shows representative examples from within each quartile of the SFG error distribution, selected randomly and color-coded to match their respective bins.

\begin{figure}
\centering
\includegraphics[width=\textwidth]{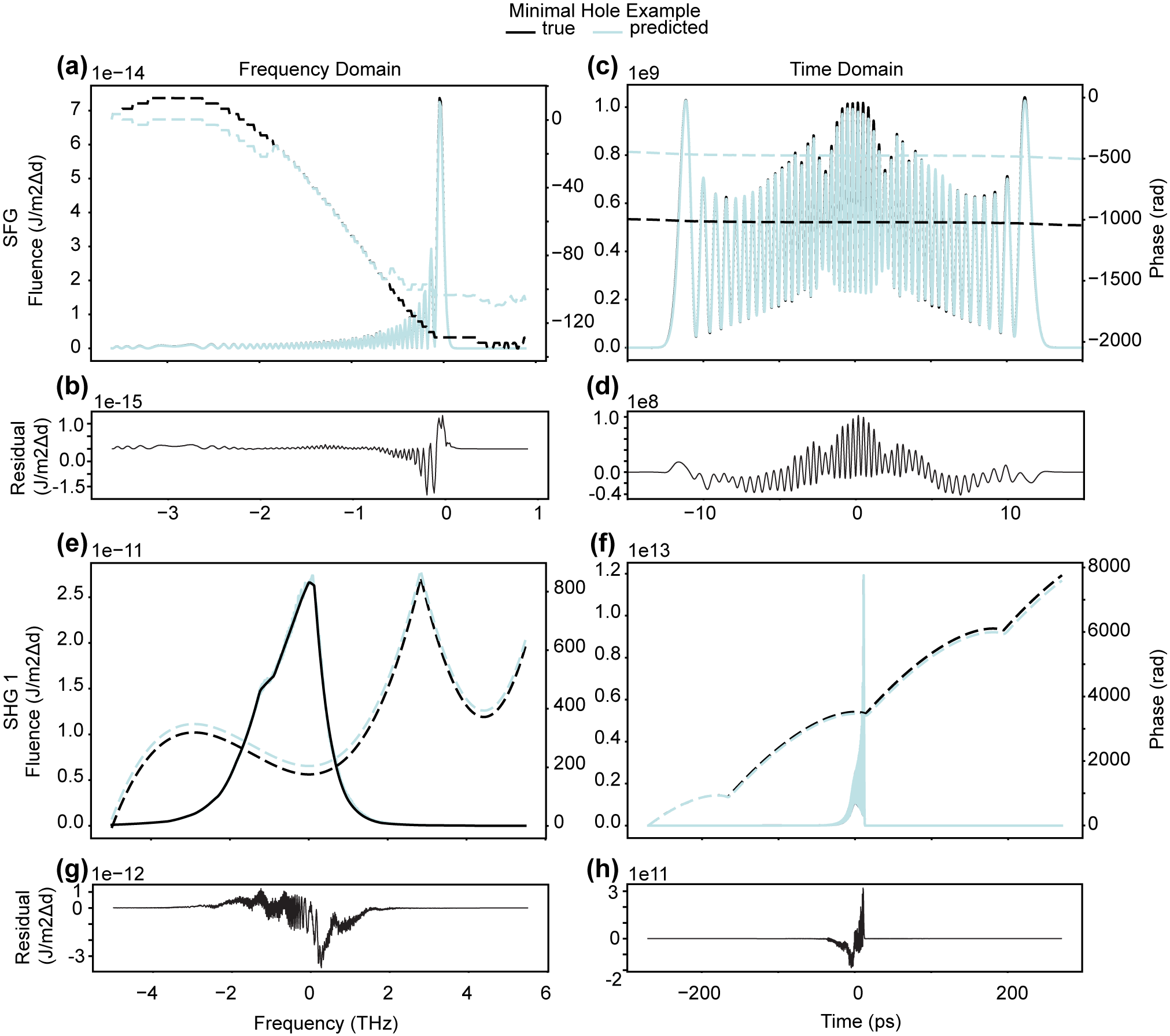}
\caption 
{ \label{fig:results1}
Minimal spectral amplitude shaping example. Example pulled from the top error quartile of SFG error distribution with primarily phase shaping. Shows frequency domain (left column) and time domain (right column) for SFG (top row) and SHG1 (bottom row) for ground truth simulation (black) versus ML inference prediction (light blue), along with the associated residuals.} 
\end{figure}

\begin{figure}
\centering
\includegraphics[width=\textwidth]{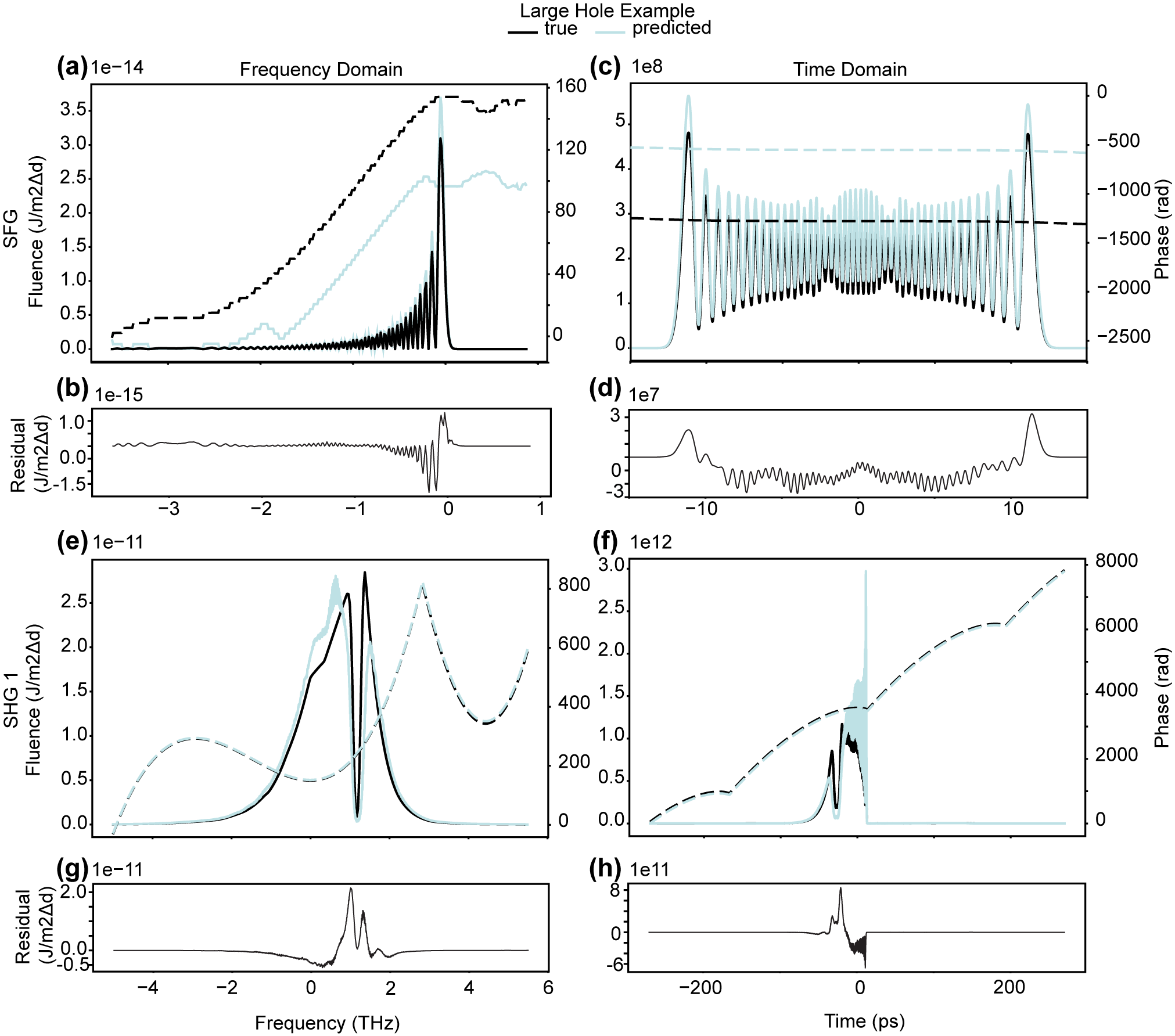}
\caption 
{ \label{fig:results2}
Large spectral amplitude shaping example. Example pulled from top error quartile of SFG error distribution with amplitude and phase shaping. Shows frequency domain (left column) and time domain (right column) for SFG (top row) and SHG1 (bottom row) for ground truth simulation (black) versus ML inference prediction (light blue) along with the associated residuals. 
} 
\end{figure}

Two additional examples from the top error distribution quartile with combined error metric values of 0.012 and 0.003, respectively, are shown in Figs.~\ref{fig:results1}~and~\ref{fig:results2} to highlight model performance across different shaping conditions. 
Figure~\ref{fig:results1} corresponds to a case with primarily phase-only shaping and minimal spectral amplitude modulation, while Fig.~\ref{fig:results2} shows a case with a pronounced spectral hole. 
In both cases, the model accurately reconstructs the SFG and SHG1 signals in both spectral and temporal domains, with only localized discrepancies observed in SHG1 when large spectral modulations are present.
These examples demonstrate that the model generalizes well across a range of spectral shaping conditions.

\subsection{Efficiency}
A central motivation for the LSTM-based surrogate model is to accelerate simulation of nonlinear $\chi^{(2)}$ processes while preserving high fidelity. 
In our start-to-end pipeline--from pulse shaping through amplification and noncollinear SFG--the nonlinear propagation step, modeled via SSFM, dominates the total runtime. 
Each instance requires approximately 1.875 seconds for SSFM alone out of the 1.98 seconds for the full simulation on a single CPU core.

Table~\ref{tab:speedUp} reports inference times for the LSTM surrogate across hardware configurations. 
On a CPU, the surrogate recovers the baseline simulation time with a slight overhead due to batching operations, resulting in a marginal increase to 2.01 seconds per instance. 
In contrast, execution on a single NVIDIA A100 GPU with a batch size of 200 reduces inference time to 7.43 milliseconds per instance, yielding a 252$\times$ speedup relative to SSFM.

\begin{table}[h!]
\centering
\caption{\label{tab:speedUp}Speed-up statistics.}
\renewcommand{\arraystretch}{1}
\begin{tabular}{|>{\raggedright\arraybackslash}m{2.6cm}|
                >{\centering\arraybackslash}m{1.8cm}|
                >{\centering\arraybackslash}m{1.5cm}|
                >{\centering\arraybackslash}m{2.8cm}|
                >{\centering\arraybackslash}m{2.5cm}|
                >{\centering\arraybackslash}m{1.8cm}|}
\hline
 & \textbf{Resources} & \textbf{Batch Size} & \textbf{Time for 1,000 Instances (s)} & \textbf{Time per Instance (s)} & \textbf{Speedup (folds)} \\ \hline
\textbf{Baseline Simulation} & 1 CPU & 1 & 1,875 & 1.8750 & -- \\ \hline
\textbf{LSTM + CPU Acceleration} & 1 CPU & 200 & 2,015.6 & 2.0156 & 0.9 \\ \hline
\textbf{LSTM + GPU Acceleration} & 1 GPU & 200 & 7.43 & 0.0074 & 252.4 \\ \hline
\end{tabular}
\end{table}

This acceleration represents a significant reduction in computational burden and demonstrates the surrogate’s ability to reproduce nonlinear field evolution orders of magnitude faster than traditional methods.

%% file: sections/discussion.tex
Accurately evaluating surrogate model performance in nonlinear optics remains nontrivial, particularly when comparing high-resolution simulated fields with diagnostics of finite resolution. 
In early experiments, we tested SSIM alongside MSE/NMSE, cosine similarity, and variants computed on the complex field, intensity envelope, and total energy. 
While SSIM is attractive for its perceptual grounding, it is primarily designed for 2D imagery and also requires nontrivial choices (windowing, cropping) to adapt to the representative 1D temporal signals.
Moreover, no single metric consistently aligned with qualitative visual assessment across the diversity of pulse shapes and energy scales in our dataset.

Motivated by these observations, we adopted a composite error metric in which lower values indicate better agreement.
This metric combines three complementary terms: (i) inverted cosine similarity to emphasize shape agreement independent of magnitude, (ii) an area-based NMSE to penalize energy mismatch, and (iii) a Wasserstein distance to capture localized redistributions in the intensity profile. 
Cosine similarity has precedent in optics as a sensitive spectral similarity measure, e.g., for optical frequency combs~\cite{irimatsugawa2021cosine}. 
Component-wise Gaussian smoothing and normalization are applied to stabilize each term, and the final score is computed via a quadratic combination of the scaled components.

A persistent challenge--independent of metric choice--is the trade-off between structural fidelity and intensity magnitude in signals with large dynamic range. 
Small discrepancies in high fluence regions can dominate error, whereas purely shape-based measures can overlook meaningful energy deviations. 
Our composite metric mitigates (but does not eliminate) this tension by explicitly allocating terms to each aspect. 
We stress that metric design is application-specific: crop windows, normalization modes, smoothing scales, clipping ranges, weights, and the combination rule are hyperparameters that should be tuned to the diagnostic, operating regime, and end-use objective.
We therefore interpret scores comparatively (e.g., distributions and quartiles over a test set) rather than prescribing universal thresholds. 
Alternatively, comparison could be performed in a learned latent space--using an auxiliary model to project signals into a feature representation where similarity is more perceptually or physically meaningful--but we leave such approaches to future work.

Practical interpretation should account for finite instrument resolution where discrepancies that are numerically detectable at simulation resolution may be immaterial experimentally. 
For instance, ultraviolet spectrometers typically resolve spectral features no finer than $\sim$0.1 nm ($\approx$0.12 THz), and temporal diagnostics such as cross-correlators often have resolution limits around 80 fs. 
Our simulations, in contrast, resolve features down to $\sim$1.2 fs across pulses that are nearly 20 ps in duration. 
As a result, the model may deviate from SSFM in ways that are quantitatively measurable in simulation but imperceptible within typical experimental resolution. 
Based on simulation alone, we interpret the error scores below 0.015 (25th percentile) as excellent, between 0.015--0.030 (50th percentile) as acceptable, between 0.030--0.067 (75th percentile) as conditionally acceptable, above 0.067 as indicative of potentially significant discrepancies--though the thresholds remain context dependent and can be relaxed when accounting for laboratory diagnostics.

An additional consideration is the choice of which output fields to evaluate. 
In this work, we focus on the SFG output ($A_2$) due to its central role in ultraviolet pulse shaping. 
The SFG pulse both starts from zero amplitude and evolves significantly through the nonlinear medium, making it more sensitive to model error. 
The SHG outputs ($A_1$ and $A_3$) evolve less dramatically, but, interestingly, we observe that in the low SFG combined error metric examples, SHG1 also matches well--even though it was not used to sort predictions. 
This potentially indicates that the LSTM surrogate captures global propagation dynamics rather than overfitting to a specific output. 
However, at the highest-error example in Fig.~\ref{fig:err_hist}d, SHG1 matches closely while the SFG intensity profiles agree in structure but differ significantly in magnitude—illustrating that the metric properly penalizes such cases and suggesting that the model may not fully capture the coupled $\chi^{(2)}$ physics. 
While training did involve a metric assessing both SFG and SHG1/2 waveforms in the frequency domain, future extensions could consider composite evaluation metrics that jointly assess multiple fields, better aligning with the coupled nature of the interaction.

In terms of model generalization, one promising direction is to expand the diversity of training inputs. 
While our dataset is based on physically realizable DCNS configurations using shaped IR pulses, augmenting the dataset with a broader range of mixing fields--such as randomly generated Fourier bases or combinations of Gaussians—may help improve robustness to experimental imperfections or unusual configurations.

Finally, a key strength of the LSTM surrogate lies in its domain efficiency. 
Traditional split-step methods like SSFM require alternating between time and frequency domains at every step of propagation to separately apply dispersion and nonlinear operators. 
This results in redundant transformations, high memory usage, and slower performance. 
The LSTM, by contrast, learns a direct mapping in the reduced frequency domain, eliminating the need for repeated FFTs or dual-domain storage. 
This simplification enables future real-time inference and tighter integration with experimental control systems, making the model not just accurate, but practically deployable.

%% file: sections/conclusions.tex
This work presents a surrogate modeling framework based on a Long Short-Term Memory network for simulating nonlinear $\chi^{(2)}$ optical processes. 
As a canonical example, we focus on noncollinear sum-frequency generation in a dispersion-controlled nonlinear synthesis process, demonstrating accurate modeling of field evolution across a wide range of pulse shapes.
Trained on high-resolution split-step Fourier method simulations, the LSTM surrogate achieves over 250$\times$ speedup while operating entirely in the reduced frequency domain--eliminating the need for repeated time-frequency domain conversions and reducing computational overhead.

While demonstrated here for SFG, the framework is broadly applicable to other $\chi^{(2)}$ processes and nonlinear geometries. 
This acceleration is particularly impactful for applications requiring rapid design iterations, adaptive laser control, and large-scale parameter space exploration--critical demands in integrated photonics and high-power laser systems. 
In the context of accelerator photocathode drive lasers, for instance, our model enables fast predictions of nonlinear pulse propagation, supporting advanced shaping and optimization strategies that were previously hindered by computational cost.

Beyond its immediate utility for accelerating nonlinear simulations, this work lays the foundation for integrating machine learning surrogates into experimental workflows. 
In particular, the approach supports the development of photonic digital twins--self-updating predictive models that couple high-fidelity simulations with real-time experimental feedback. 
While this study focuses on simulation-trained surrogates, the framework naturally extends to hybrid modeling strategies that incorporate empirical data, enabling inverse design, adaptive control, and physics-informed generalization across evolving conditions.

Ultimately, this work demonstrates how domain-specific machine learning models can significantly enhance the speed, flexibility, and reach of nonlinear optical modeling.
By bridging the gap between computational simulations and experimental implementation, it opens new avenues for efficient, scalable, and intelligent photonic system design.